\documentclass[journal]{IEEEtran}

\usepackage{graphicx}
\usepackage{cite}
\usepackage{multirow}
\usepackage[cmex10]{amsmath}
\usepackage{hyperref}
\usepackage{algorithm,algpseudocode}
\usepackage{algorithmicx}
\algnewcommand{\Inputs}[1]{%
  \State \textbf{Inputs:}
  \Statex \hspace*{\algorithmicindent}\parbox[t]{.8\linewidth}{\raggedright #1}
}
\algnewcommand{\Initialize}[1]{%
  \State \textbf{Initialize:}
  \Statex \hspace*{\algorithmicindent}\parbox[t]{.8\linewidth}{\raggedright #1}
}

\begin{document}
%
\title{Fast and Lightweight Rate Control for Onboard Predictive Coding of Hyperspectral Images}
%
%
%

\author{Diego~Valsesia,
        and~Enrico~Magli
\thanks{This work was supported by the European Space Agency (ESA-ESTEC)
under grant 107104.The authors are with Politecnico di Torino, Torino, Italy.}}

%
%

\markboth{IEEE GEOSCIENCE AND REMOTE SENSING LETTERS,}%
{Valsesia \MakeLowercase{\textit{et al.}}: Fast and Lightweight Rate Control for Onboard Predictive Coding of Hyperspectral Images}
%



\maketitle

\begin{abstract}

Predictive coding is attractive for compression of hyperspecral images onboard of spacecrafts in light of the excellent rate-distortion performance and low complexity of recent schemes. In this letter we propose a rate control algorithm and integrate it in a lossy extension to the CCSDS-123 lossless compression recommendation. The proposed rate algorithm overhauls our previous scheme by being orders of magnitude faster and simpler to implement, while still providing the same accuracy in terms of output rate and comparable or better image quality. 

\end{abstract}


%
\IEEEpeerreviewmaketitle

\vspace*{-0.3cm}
\section{Introduction}
Hyperspectral imaging from space-borne spectrometers has a multitude of applications, such as terrain analysis, material identification, military surveillance, etc. The ever growing spectral and spatial resolution of such instruments poses challenges in handling such wealth of information. In particular, onboard compression is of paramount importance to overcome the limited downlink bandwidth. This is a very active area of research as it poses peculiar challenges, not encountered elsewhere. In fact, onboard compression algorithms face strict complexity limitations due to constraints on the payload hardware. Several solutions based on different techniques have been proposed, such as low-complexity spatial \cite{ccsds122} and spectral transforms \cite{pot}, distributed source coding \cite{abrardoDSC}, compressed sensing \cite{universal,barducci2014compressive}, and predictive coding \cite{rizzolow,acap,calibrationartifacts}. Predictive coding is one of the most popular solutions, as it enables low-complexity, high-throughput solutions, and excellent rate-distortion performance. One of the most recent recommendations by the Consultative Committee for Space Data Systems (CCSDS) is a lossless compression algorithm based on predictive coding \cite{ccsds123} and an extension to lossy compression is ongoing. Lossy predictive coding is typically operated in the so-called near-lossless mode, where the maximum absolute error on the reconstructed pixels is bounded by a constant. However, near-lossless compression has some drawbacks, the most notable being a variable output rate dependent on the image content. In our previous works \cite{hydra,valsesia_icip} we showed that it is possible to perform simultaneous rate and quality control with a complexity compatible with resources available onboard of spacecrafts. In this letter, we propose a significantly improved rate control algorithm with respect to \cite{hydra}. The main contribution is that the new algorithm is extremely fast and lightweight with respect to \cite{hydra} and this allows mainly to \textit{i}) save hardware resources that could be dedicated to other purposes (or not used thus saving power), compared e.g. with the solution in \cite{valsesia_icip} which requires a dedicated board, running in parallel to the compressor; \textit{ii}) enable very high throughput implementations, while the original algorithm may have caused bottlenecks in such scenarios because of some complex rate-distortion optimization operations. We test the proposed algorithm by extending the CCSDS-123 recommendation, but the method is general and can be applied to any predictive encoder. 

\vspace*{-0.3cm}
\section{Related work}
\label{sec:background}

This section reviews the rate control algorithm originally proposed in \cite{hydra} and further developed in \cite{valsesia_icip}. The goal of the algorithm is to control the output rate of a predictive encoder of hyperspectral and multispectral images, under low complexity and memory constraints. To be best of our knowledge this is the only work on rate control for onboard predictive coding. However rate control algorithms are often used either onground \cite{spiht} or onboard \cite{satellite_data_compression} in transform coding schemes. The algorithm in \cite{hydra} selects quantizers operating on the prediction errors in predefined spatial and spectral regions. The rate control algorithm works on a slice-by-slice basis, where we call ``slice'' a predefined number of lines with all their spectral channels. Each band of each slice is divided into nonoverlapping $16 \times 16$ blocks. The rate control algorithm assigns a quantization step size for each of the blocks in each spectral band in order to meet the target rate with the lowest distortion. This is done via a two-stage process consisting of:
\begin{itemize}
    \item Training stage: unquantized prediction residuals in each block are modeled as realizations of independent and identically distributed Laplacian random variables. The sample variance is estimated in order to predict the rate as function of the quantization step size. The rate function is stored in a lookup table (LUT) as it is too complex to implement directly.
    \item Optimization stage: a greedy optimization algorithm is employed to select the quantization step sizes that allow achieving the target rate with the lowest distortion.
\end{itemize}
The training stage requires to run the predictor twice for each slice, once to estimate the variance of the unquantized prediction residuals, then one more time in order to perform the actual coding which also quantizes the residuals with the step sizes computed by the rate controller. This dual prediction stage can be performed in series to the coding process, as in the original algorithm, or in parallel, as in \cite{valsesia_icip}, by having an independent module that works ahead of the coding process. 

Furthermore, the algorithm measures the actual rate produced by encoding the slice with the computed quantization step sizes and uses this information to update the target rate for the next slices. This mode of operation has been shown to effectively correct inaccuracies in the model without reducing the rate-distortion performance.

In \cite{conoscenti} the authors simplify the rate control algorithm by eliminating the rate-distortion optimization phase and working on line-by-line basis instead of blocks. This idea is also used in this work, which however solves major drawbacks of the method proposed in \cite{conoscenti}. The method proposed in this letter improves over \cite{hydra},\cite{valsesia_icip},\cite{conoscenti} by introducing the following novel points:
\begin{itemize}
    \item choosing just one quantization step size per spectral line, which avoids complex rate-distortion optimization algorithms (also present in \cite{conoscenti});
    \item on-the-fly estimation of the residual statistics, which avoids running the predictor twice, with significant gains in terms of speed;
    \item simpler arithmetic that does not involve squaring operations to compute the statistical parameters;
    \item simpler arithmetic speeding up access to the rate LUT;
    \item reduced number of LUT lookups.
\end{itemize}

\vspace*{-0.2cm}
\section{Proposed method}\label{sec:proposed}

\begin{figure}
    \centering
    \includegraphics[width=0.81\columnwidth]{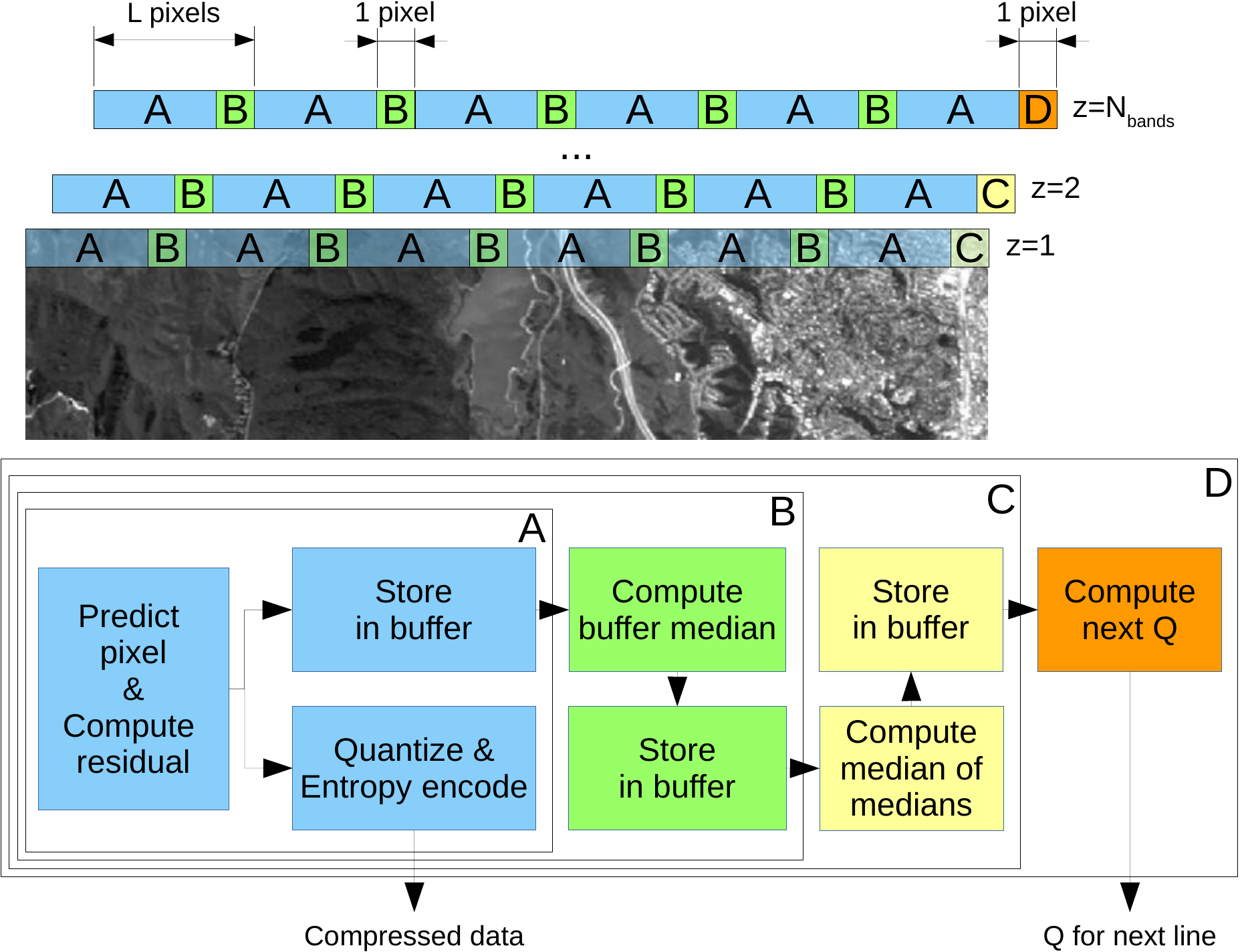}
    \vspace*{-0.1cm}
    \caption{High-level diagram of proposed method. A different number of operations are performed depending on the current pixel position. Pixel in position D (last pixel of the current line in the last band) performs all the operations of pixels A, B or C and runs the rate control procedure to determine the quantization step size for the next line.}
    \label{fig:diagram}
    \vspace*{-0.35cm}
\end{figure}

In this section we explain the main novel contributions of this work. The goal is to design a lightweight rate control algorithm that is capable of providing accurate control of the output rate of a predictive encoder while being, at the same time, simple to implement in hardware, and fast in order to enable high-throughput implementations. Fig. \ref{fig:diagram} shows a high-level diagram of the proposed method and its interaction with a predictive encoder. The main operations are described in the following subsections. In the following the term \emph{spectral line} refers to a row of pixels with all its spectral channels. It can be noticed that the algorithm proceeds one spectral line at a time and that different operations are performed depending on the spatial location of the pixel under coding. For the sake of simplicity, the following description assumes that the pixels are read and coded in band-interleaved-by-line (BIL) order. The objective of the algorithm is to encode the current spectral line, thus producing entropy-coded versions of quantized prediction residuals and to decide the quantization step size to be used to encode next spectral line. A single quantization step size $Q$ is used for each spectral line.

\vspace*{-0.2cm}
\subsection{Prediction and parameter estimation from residuals}

The use of a single quantization step size $Q$ of a uniform scalar quantizer for each spectral line is at the core of the algorithm. It is necessary to have a statistical characterization of the prediction residuals in each band in order to choose the value of $Q$ yielding the desired target rate. The statistics of prediction residuals of adjacent spectral lines are highly correlated and therefore it is possible to use the statistics of the current spectral line as a prediction for the statistics of next line. We assume that the unquantized prediction residuals are realizations of i.i.d. Laplacian random variables. The Laplacian distribution ($f_R(x) = \frac{\Lambda}{2}e^{-\Lambda\vert x \vert}$) is characterized by a single parameter $\Lambda$ .
Notice that the $\Lambda$ parameter is also related to the variance $\sigma^2$ through $\Lambda = \sqrt{\frac{2}{\sigma^2}}$.
The magnitude of a Laplace random variable follows an exponential distribution of parameter $\Lambda$, and the sample mean is the unbiased maximum likelihood estimator of $\Lambda^{-1}$. However, this is not a robust estimator \cite{huber2011robust}. Replacing it with the sample median has several advantages, including robustness to outliers (the residuals are far from perfectly i.i.d. Laplacian), and simpler arithmetic (avoids division by the length $N_\mathrm{cols}$ of the line). Computing the exact median of the prediction residuals of an entire row would require a sorting operation of complexity $\mathcal{O}(N_\mathrm{cols}\log N_\mathrm{cols})$. A good approximation can be obtained with the median of medians algorithm \cite{medians} which breaks down a row into non-overlapping subsets of $L$ contiguous residuals. The median of each subset is computed and then the median of medians provides an approximation of the true median of the row. This approach reduces the complexity to $\mathcal{O}\left( N_\mathrm{cols}\log L + \frac{N_\mathrm{cols}}{L} \log \frac{N_\mathrm{cols}}{L} \right)$.
The parameter $L$ is set as a compromise of several objectives. In fact, $L$ should be large enough that the approximation to the median by means of the median of median is satisfactory. However, a large $L$ increases the complexity of the calculation. A good design should approximately balance the complexity of the local median operations (each costing $\mathcal{O}(L \log L)$) and the complexity of the final median performed at the end of the row (costing $\mathcal{O}(\frac{N_{cols}}{L} \log \frac{N_{cols}}{L})$) for the typical width of an image, so that throughput is maximized by the absence of bottlenecks.

 The coding and estimation jobs proceed together. Following the notation of Fig. \ref{fig:diagram}, pixels of type A entropy-code the current quantized prediction residual of the current pixel but also store in a temporary buffer the unquantized residual. The encoder reaches a type-B pixel after $L-1$ type-A pixels and the temporary buffer holds the unquantized prediction residuals for that subset of type-A pixels. After the computation of the unquantized residual for the type-B pixel, the median of the buffer is computed and stored in the buffer of medians. This sequence of $L-1$ pixels of type A terminated by a type B pixel is repeated until the end of the row. The row is terminated by a pixel of type C, which computes both the median for the local subset of pixels and the median of medians. Each band $z$ stores its own median of medians $m_z$. The last pixel in the last band is of type D and it uses all the medians of medians to run the procedure to determine the quantization step size to be used for the next spectral line.

There are several important differences with the algorithms in \cite{hydra,conoscenti}. In particular, a major shortcoming of those algorithms is the need to run the predictor twice, once for estimation of the statistics and once for the actual coding. Moreover, they adopt a more computationally expensive approach to parameter estimation by computing the sample variance $\sigma^2$ of the unquantized prediction residuals, and then using it to compute $\Lambda$. This process requires some computational effort when performed with integer arithmetic and is subject to numerical approximations.

\vspace*{-0.2cm}
\subsection{Assignment of quantization step size}

\setlength{\textfloatsep}{0pt}
\begin{algorithm}
\begin{algorithmic}
\Inputs{$R_\mathrm{target}, L, Q^\mathrm{max}$}
\Initialize{$Q_1=1$}
\State
\For{$y = 1, \ldots, N_\mathrm{rows}$}
    \For{$z = 1, \ldots, N_\mathrm{bands}$}
        \State $i \gets 0$
        \For{$x = 1, \ldots, N_\mathrm{cols}$}
            \State  $\hat{s}_{x,y,z} \gets$ prediction of pixel $s_{x,y,z}$ 
            \State Residual $r_{x,y,z} \gets \hat{s}_{x,y,z} - s_{x,y,z}$
            \If{$x \mod L = L-1$}
                \State $\tilde{m}_i \gets $ median($r_{x,y,z},\ldots,r_{x-L+1,y,z}$)
                \State $i\gets i+1$
            \EndIf
            \State Quantized $\hat{r}_{x,y,z} \gets \mathrm{sgn}(r_{x,y,z}) \cdot \left\lfloor \frac{\vert r_{x,y,z} \vert + \frac{Q-1}{2}}{Q_y} \right\rfloor$
        \EndFor
        \State $m_z \gets $ median($\tilde{m}_0,\ldots,\tilde{m}_{(N_\mathrm{rows}/L)}$)
    \EndFor
    \State Measure actual output rate
    \State Update $R_\mathrm{target}$ using Eq. (14) from \cite{hydra}
    \State $Q_{y+1} \gets Q_{y}$
    \State $R \gets \sum_{z=1}^{N_\mathrm{bands}} R(m_z,Q_{y+1})$
    \If{$R \geq R_\mathrm{target}$}
        \While{$R \geq R_\mathrm{target}$ AND $Q_{y+1}<Q^\mathrm{max}$}
            \State $R_\mathrm{old} \gets R$
            \State $Q_{y+1} \gets Q_{y+1} + 2$
            \State $R \gets \sum_{z=1}^{N_\mathrm{bands}} R(m_z,Q_{y+1})$
        \EndWhile
        \If{$\vert R - R_\mathrm{target} \vert > \vert R_\mathrm{old} - R_\mathrm{target} \vert$}
            \State $Q_{y+1} \gets Q_{y+1} - 2$
        \EndIf
    \Else
        \While{$R \leq R_\mathrm{target}$ AND $Q_{y+1}>1$}
            \State $R_\mathrm{old} \gets R$
            \State $Q_{y+1} \gets Q_{y+1} - 2$
            \State $R \gets \sum_{z=1}^{N_\mathrm{bands}} R(m_z,Q_{y+1})$
        \EndWhile
        \If{$\vert R - R_\mathrm{target} \vert > \vert R_\mathrm{old} - R_\mathrm{target} \vert$}
            \State $Q_{y+1} \gets Q_{y+1} + 2$
        \EndIf
    \EndIf

\EndFor
\end{algorithmic}
\caption{Coding and Rate Control}
\label{alg:ratecontrol}
\end{algorithm}
\setlength{\textfloatsep}{8pt}

The procedure that assigns the quantization step sizes is extremely lightweight in order to provide the fastest implementation possible. It takes as input the set $\lbrace m_z \rbrace_{z=1}^{N_\mathrm{bands}}$ of the medians of medians for all the bands. The procedure uses a rate LUT to find the odd integer-valued quantization step size $1 \leq Q \leq Q^{\mathrm{max}}$ such that:
\begin{align*}
    Q = \arg\min_{q \in \lbrace 1,3,5,\dots,Q^\mathrm{max} \rbrace } \left\vert \sum_{z=1}^{N_\mathrm{bands}} R(m_z,q) - R_{\mathrm{target}} \right\vert   ,
\end{align*}
i.e., the one providing a total rate for the line and all its spectral channels closest to the target rate $R_{\mathrm{target}}$.
The rate LUT implements the following formula (from \cite{hydra}), which is the entropy of an i.i.d. Laplacian source quantized with a uniform scalar quantizer of step size $Q$ and parameter $\Lambda = \frac{1}{m}$.
\begin{align}
\label{eq:rate}
    &R(m,Q) = -\left( 1 - e^{-\frac{Q}{2m}} \right) \log_2 \left( 1-e^{-\frac{Q}{2m}} \right) + \nonumber\\
    &- \frac{e^{-\frac{Q}{2m}}}{\log(2)} \left[\log\left( \frac{1-e^{-\frac{Q}{m}}}{2} \right) +\frac{Q}{2m} - \frac{Q}{m\left( 1-e^{-\frac{Q}{m} } \right) } \right]
\end{align}
The proposed method uses two variables, namely $m$ and $\delta=(Q-1)/2$, to index the rate LUT that is treated as a two-dimensional matrix storing the values of $R(m,Q)$ precomputed using Eq.\eqref{eq:rate}. Notice that using $\delta$ instead of $Q$ as index for the LUT is slightly more efficient (one fewer bit of dynamic range). This technique is faster than the one used in \cite{hydra} which uses a single-variable indexing. Indeed the rate depends only on the product $\Lambda Q$ and the parameter estimation algorithm of \cite{hydra} directly returns $\Lambda$. However, computing the correct position in the table in such manner is inefficient because it requires to perform a multiplication every time $Q$ or $\Lambda$ are changed. Moreover, the rate function was sampled at predefined values of $\Lambda Q$ and a suitable mapping of the product $\Lambda Q$ provided at runtime with the closest in the LUT had to be computed. The proposed method overcomes such limitations by exploiting the fact that $m$ is discrete-valued and with limited dynamic range as well as not requiring any index calculation.

A further optimization is included to minimize the number of LUT lookups. The procedure that determines the quantization step size for the following spectral line uses the last chosen quantization step size as a prediction for the next one. Then it checks if such quantization step size yields a rate above or below the target, consequently deciding whether to increase it or decrease it. Since the quantization step sizes used in successive spectral lines are expected to be close, this strategy minimizes the expected number of times the memory storing the LUT is queried to compute $R(m,Q)$, thus reducing latency.

As a final remark, in a tradeoff between efficiency and available features, some characteristics of the algorithm in \cite{hydra} have been traded for increased speed. In particular, the original algorithm performed simultaneous control of rate and quality on image blocks, thus enabling the definition of spatial as well as spectral regions of interest with different rate-distortion policies. This feature is partially lost in the faster version as a single quantization step size is used for all the bands of a line.

The full algorithm, including both estimation of the statistics of residuals and computation of the quantization step sizes, is reported in Algorithm \ref{alg:ratecontrol}. Notice that the initialization of $Q_1$ may be different from $Q_1=1$ and, depending on the particular target rate, a higher $Q$ might be appropriate.

\setlength{\textfloatsep}{10pt}

\vspace*{-0.2cm}
\section{Experimental Results}


\begin{table}
\caption{Test Images}
\label{table:images}
\centering
\begin{tabular}{c c c c c c c}
\multirow{2}{*}{Image} & \multirow{2}{*}{Rows} & \multirow{2}{*}{Columns} & \multirow{2}{*}{Bands} & \multirow{2}{*}{$P$} & Bit & Entropy \\
 &  &  &  &  & depth & (bpp) \\
\hline
\hline
\textsc{AVIRIS sc0\_raw} & 512 & 680 & 224 & 15 & 16 & 12.62\\
\hline
\textsc{AIRS gran9} & 135 & 90 & 1501 & 10 & 14 & 11.16\\
\hline
\textsc{CASI-t0477f06-nuc} & 1225 & 406 & 72 & 2 & 16 & 10.65\\ 
\hline
\textsc{CRISM-sc167-nuc} & 510 & 640 & 545 & 3 & 16 & 10.54\\ 
\hline
\textsc{LANDSAT mountain} & 1024 & 1024 & 6 & 5 & 8 & 6.33\\ 
\hline
\textsc{MODIS-MOD01day} & 2030 & 1354 & 14 & 2 & 12 & 7.99\\ 
\hline
\end{tabular}
\vspace*{-0.1cm}
\end{table}

\begin{table}
\caption{Accuracy of rate control}
\label{table:accuracy}
\centering
\begin{tabular}{c c c c c c c}
Image & Method & 0.5 bpp & 1 bpp & 2 bpp & 3 bpp & 4 bpp \\
\hline
\hline
\textsc{AVIRIS} & Proposed & \bf 0.501 & \bf 1.002 & 2.002 & \bf 3.001 & \bf 4.001 \\[0.01cm]
\textsc{sc0\_raw} & \cite{hydra} & 0.510 & 1.005 & \bf 2.000 & 2.990 & 3.998\\
\hline
\textsc{AIRS} & Proposed & \bf 0.502 & \bf 1.006 & \bf 2.001 & \bf 2.993 & \bf 4.021 \\[0.01cm]
\textsc{gran9}& \cite{hydra} & 0.487 & 0.987 & 1.967 & 2.942 & 3.930\\
\hline
\textsc{CASI} & Proposed & \bf 0.501 & \bf 1.001 & \bf 2.001 & \bf 3.001 & \bf 4.000\\[0.01cm] 
\textsc{t0477f06-nuc}& \cite{hydra} & 0.402 & 1.142 & 2.011 & 3.007 & 4.007\\
\hline
\textsc{CRISM} & Proposed & \bf 0.501 & \bf 1.003 & \bf 2.004 & \bf 3.001 & \bf 4.002\\[0.01cm] 
\textsc{sc167-nuc}& \cite{hydra} & 0.508 & \bf 1.003 & 1.993 & 2.986 & 3.992\\
\hline
\textsc{LANDSAT} & Proposed & \bf 0.502 & \bf 1.001 & \bf 2.001 & \bf 3.002 & 3.736$^{(*)}$\\[0.01cm] 
\textsc{mountain}& \cite{hydra} & 0.435 & 0.898 & 2.008 & 2.838 & 3.736$^{(*)}$\\
\hline
\textsc{MODIS} & Proposed & \bf 0.508 & \bf 1.002 &  \bf 2.005 & \bf 3.006 & \bf 4.005\\[0.01cm] 
\textsc{MOD01day} & \cite{hydra} & 0.478 & 1.018 & 2.016 & 3.008 & 4.006\\
\hline
\end{tabular}\\
\vspace*{0.1cm}
(*): lossless
\vspace*{-0.3cm}
\end{table}

\begin{table}
\caption{Image Quality}
\label{table:snr}
\centering
\begin{tabular}{c c c c c c}
\multirow{2}{*}{Image} & \multirow{2}{*}{Rate} & \multicolumn{2}{c}{Proposed} & \multicolumn{2}{c}{\cite{hydra}}\\
& & SNR (dB) & MAD & SNR (dB) & MAD\\
\hline
\hline
\multirow{5}{*}{\textsc{AVIRIS sc0\_raw}} & 0.5 bpp & 39.09 & \bf 255 & \bf 39.99 & \bf 255 \\
& 1 bpp & 44.60 & 255 & \bf 47.50 & \bf 228 \\
& 2 bpp & 57.79 & \bf 25 & \bf 57.90 & \bf 25\\
& 3 bpp & \bf 64.10 & \bf 25 & 63.27 & \bf 25 \\
& 4 bpp & \bf 71.13 & \bf 3 & 71.07 & \bf 3\\
\hline
\multirow{5}{*}{\textsc{AIRS gran9}} & 0.5 bpp & 38.12 & 255 & \bf 46.41 & \bf 53 \\
& 1 bpp & 53.86 & 18 & \bf 54.93 & \bf 10 \\
& 2 bpp & 63.04 & \bf 4 & \bf 63.31 & \bf 4\\
& 3 bpp & 67.28 & 3 & \bf 69.64 & \bf 1\\
& 4 bpp & \bf 79.25 & 1 & 76.78 & \bf 1\\
\hline
\multirow{5}{*}{\textsc{CASI-t0477f06-nuc}} & 0.5 bpp & \bf 32.37 & \bf 255 & 27.32 & \bf 255 \\
& 1 bpp & \bf 41.69 & \bf 49 & 38.12 & 77 \\
& 2 bpp & \bf 50.92 & \bf 7 & 49.45 & 13\\
& 3 bpp & \bf 57.64 & \bf 4 & 56.51 & 8\\
& 4 bpp & 62.03 & \bf 3 & \bf 62.89 & \bf 3\\
\hline
\multirow{5}{*}{\textsc{CRISM-sc167-nuc}} & 0.5 bpp & \bf 37.53 & 144 & 36.95 & \bf 84 \\
& 1 bpp & 44.14 & 48 & \bf 44.17 & \bf 27\\
& 2 bpp & 52.59 & \bf 7 & \bf 52.67 & 9\\
& 3 bpp & \bf 59.37 & \bf 3 & 58.43 & 4\\
& 4 bpp & 64.25 & \bf 2 & \bf 64.67 & 4\\
\hline
\multirow{5}{*}{\textsc{LANDSAT mountain}} & 0.5 bpp & \bf 21.06 & 41 & 19.90 & \bf 35 \\
& 1 bpp & \bf 27.13 & \bf 10 & 22.88 & 58\\
& 2 bpp & \bf 34.17 & \bf 3 & 32.94 & 5\\
& 3 bpp & \bf 39.37 & \bf 3 & 38.49 & 4\\
& 4 bpp & $\mathbf{\infty}$ & \bf 0 & $\mathbf{\infty}$ & \bf 0 \\
\hline
\multirow{5}{*}{\textsc{MODIS-MOD01day}} & 0.5 bpp & \bf 30.72 & \bf 255 & 29.51 & 255 \\
& 1 bpp & \bf 38.70 & 230 & 37.38 & \bf 214\\
& 2 bpp & \bf 49.59 & 134 & 49.12 & \bf 68\\
& 3 bpp & \bf 58.66 & \bf 17 & 58.22 & 28\\
& 4 bpp & 63.90 & 70 & \bf 66.09 & \bf 10\\
\hline
\end{tabular}
\vspace*{-0.3cm}
\end{table}

\begin{table*}
\caption{Average LUT lookups per megapixel and rate control runtime (sec.)}
\label{table:lookups}
\centering
\begin{tabular}{c c c c c c c c c c c c}
\multirow{2}{*}{Image} & \multirow{2}{*}{Algorithm} & \multicolumn{2}{c}{0.5 bpp} & \multicolumn{2}{c}{1 bpp} &  \multicolumn{2}{c}{2 bpp} &  \multicolumn{2}{c}{3 bpp} &  \multicolumn{2}{c}{4 bpp} \\
\cline{3-12}
& & Lookups & Time & Lookups & Time & Lookups & Time & Lookups & Time & Lookups & Time \\
\hline
\hline
\multirow{2}{*}{\textsc{AVIRIS sc0\_raw}} & Proposed & \bf 13,485 & \bf 0.0031 & \bf 6,250 & \bf 0.0026 & \bf 3,412 & \bf 0.0016 & \bf 3,349 & \bf 0.0019 & \bf 3,553 & \bf 0.0017 \\
& \cite{hydra} & 615,567 & 11.8774 & 371,675 & 7.4340 & 257,333 & 5.4265 & 235,065 & 5.0735 & 240,722 & 5.9426\\
\hline
\multirow{2}{*}{\textsc{AIRS gran9}} & Proposed & \bf 79,506 & \bf 0.0024 & \bf 24,198 & \bf 0.0008 & \bf 23,704 & \bf 0.0008 & \bf 24,115 & \bf 0.0008 & \bf 22,058 & \bf 0.0008\\
& \cite{hydra} & 369,916 & 1.5309 & 288,757 & 1.3689 & 273,748 & 1.5250 & 276,991 & 1.6256 & 242,905 & 1.0447\\
\hline
\multirow{2}{*}{\textsc{CASI-t0477f06-nuc}} & Proposed & \bf 8,149 & \bf 0.0015 & \bf 5,503 & \bf 0.0015 & \bf 5,125 & \bf 0.0015 & \bf 5,626 & \bf 0.0014 & \bf 5,001 & \bf 0.0014 \\
& \cite{hydra} & 767,673 & 6.7921 & 738,115 & 4.6862 & 576,056 & 3.7174 & 386,564 & 2.5567 & 362,650 & 2.4206\\
\hline
\multirow{2}{*}{\textsc{CRISM-sc167-nuc}} & Proposed & \bf 5,863 & \bf 0.0022 & \bf 3,713 & \bf 0.0017 & \bf 3,321 & \bf 0.0015 & \bf 3,398 & \bf 0.0013 & \bf 3,165 & \bf 0.0015 \\
& \cite{hydra} & 389,630 & 33.4665 & 282,061 & 20.1007 & 237,631 & 16.1764 & 234,546 & 16.8475 & 240,384 & 17.0099\\
\hline
\multirow{2}{*}{\textsc{LANDSAT mountain}} & Proposed & \bf 2,418 & \bf 0.0007 & \bf 2,175 & \bf 0.0009 & \bf 2,244 & \bf 0.0009 & \bf 1,957 & \bf 0.0009 & \bf 1,951 & \bf 0.0007 \\
& \cite{hydra} & 382,011 & 0.4723 & 274,881 & 0.3482 & 246,634 & 0.3190 & 234,428 & 0.2991 & 209,264 & 0.2672\\
\hline
\multirow{2}{*}{\textsc{MODIS-MOD01day}} & Proposed & \bf 5,559 & \bf 0.0024 & \bf 5,155 & \bf 0.0023 & \bf 3,445 & \bf 0.0020 & \bf 1,729 & \bf 0.0019 & \bf 1,893 & \bf 0.0019 \\
& \cite{hydra} & 1,461,049 & 11.8816 & 1,179,269 & 8.2141 & 990,538 & 6.8846 & 1,337,818 & 9.3214 & 606,623 & 3.9860\\
\hline
\end{tabular}
\vspace*{-0.3cm}
\end{table*}

In this section we study the performance of the proposed rate control method. The algorithm is integrated into a lossy extension of CCSDS-123 where the prediction residuals are quantized with a uniform scalar quantizer and entropy coded with a range encoder \cite{martinrange} (refer to \cite{hydra} for more details). Since the focus of this paper is on the rate control algorithm we omit comparisons with other compression algorithms (e.g. M-CALIC \cite{mcalic}) which do not provide rate control on a predictive encoder. We refer the reader to \cite{tutorial} for comparisons between M-CALIC and CCSDS-123 in near-lossless and rate-controlled (using \cite{hydra}) mode, where it is shown that the latter typically outperforms the former.  All tests are conducted with the feedback version of the rate control algorithm, with the rate adjusted using Eq. (14) from \cite{hydra} (parameters $\tau=5$, $\vert \mathcal{I} \vert = 1$). The rate LUT stores the value of $R(m,Q)$ multiplied by 1000 as an integer for the following domains of the input variables: $m \in [0,1023]$, $Q=\lbrace 1,3,5,\dots, 511 \rbrace$. The choice of $Q^{max}=511$ allows the algorithm to reach very low rates (if needed according to the statistics of the residuals and the target rate). Choosing a lower value of $Q^{max}$ is possible but should not conflict with the target rate otherwise the rate controller will be unable to reach the desired rate. A lower value that is compatible with the target rate has the effect of a constraint on the maximum error and can often improve quality, as shown in \cite{hydra} and in Table \ref{table:qmax}, by forbidding excessive quantization in presence of inaccurate statistics of the residuals. The LUT uses 16-bit integers to represent the rate values, thus requiring 512 kB of storage in a software implementation. Notice that this requirement could be reduced by subsampling the available quantization step sizes (trading some accuracy in the rate) or choosing a $Q^{max}$ lower than 511 (if the expected target rate is high). The median estimation uses subsets of length $L=17$ pixels. This value was chosen to approximately equalize the complexity of the median computed at type-B pixels and the median computed at type-C pixels as explained in Sec. \ref{sec:proposed}. However, results in terms of image quality and rate are not very sensitive to the choice of this parameter. 

All the following tests were performed on images extracted from the corpus\footnote{Available at \url{http://cwe.ccsds.org/sls/docs/sls-dc/123.0-B-Info/TestData}} defined by the Multispectral and Hyperspectral Data Compression (MHDC) Working Group of the CCSDS for performance evaluation and testing of compression algorithms. We will use an ultraspectral image from the AIRS sensor, hyperspectral images from the AVIRIS, CRISM and CASI sensors and multispectral images from Landsat and MODIS sensors. Table \ref{table:images} reports the test images and their dimensions, bit-depth and entropy, as well as the number of bands $P$ used by the predictor.

The first test is concerned with the accuracy of the rate control algorithm in terms of output rate. Table \ref{table:accuracy} reports the output rate for various target rates. As can be seen, the proposed algorithm achieves remarkable accuracy for all target rates and for both the hyperspectral and multispectral cases. 

Table \ref{table:snr} shows a comparison with the algorithm in \cite{hydra} in terms of SNR defined as
\begin{align*}
    \mathrm{SNR} = 10\log_{10} \frac{\sum_{i=1}^{N_\mathrm{pixel}} s_i^2}{\sum_{i=1}^{N_\mathrm{pixel}}(s_i-\tilde{s}_i)^2}
\end{align*}
being $s_i$ and $\tilde{s}_i$ the $i$-th pixel in the original image and in the decoded image, respectively, and in terms of Maximum Absolute Distortion (MAD), the largest deviation of a reconstructed pixel from the original one.
It can be noticed that the image quality is comparable with the one produced by the original algorithm, which provided a significant improvement over some low-complexity transform coding approaches \cite{pot}. Remarkably, the proposed algorithm sometimes even outperforms the reference despite its lower complexity.

In order to test the increased efficiency of the algorithm we measured the total number of lookups to the rate LUT needed by the rate control algorithm and compared it with the algorithm in \cite{hydra}. Table \ref{table:lookups} shows the number of lookups normalized by the image size in millions of pixels. Notice that the proposed algorithm improves over the old one by reducing the number of lookups by roughly two orders of magnitude. Notice that the number of lookups is directly related to the total number of operations performed by the rate controller, implying that a speedup of at least two orders of magnitude is expected. In order to confirm this we also measured the total execution time for the portion of code responsible for rate control, also reported in Table \ref{table:lookups}, under the Time column. The results confirm that the proposed algorithm is several orders of magnitude faster, needing milliseconds when the original required seconds to perform its operations.

\begin{table}
\caption{Impact of maximum error constraint}
\label{table:qmax}
\centering
\begin{tabular}{c c c c c}
Image & $Q^{max}$ & Rate (bpp) & SNR (dB) & MAD\\
\hline
\hline
 & 511 & 1.002 & 44.60 & 255\\
\textsc{AVIRIS sc0\_raw} & 127 & 1.002 & 49.26 & 63\\
\textsc{1 bpp target} & 63 & 1.001 & 50.36 & 31\\
 & 11 & 2.656 & 63.25 & 5\\
\hline
\end{tabular}
\vspace*{-0.2cm}
\end{table}

\vspace*{-0.3cm}
\section{Conclusion}

We proposed a novel rate control algorithm for predictive coding of hyperspectral images onboard of spacecrafts and tested it by integrating it into a lossy extension of the CCSDS-123 recommendation. The proposed algorithm loses some flexibility with respect to \cite{hydra} in terms of spatial modulation of quantization step sizes, but it is orders of magnitude faster enabling simpler and higher throughput implementations, while providing comparable or better image quality.

\vspace*{-0.1cm}

\end{document}